\documentclass{emulateapj}
\usepackage{graphicx}
\usepackage{epsfig}
\def\markcite#1{} 

\def\Msun{{\rm M}_\odot}

\begin{document}
\title{The \ion{H}{2} Region of a Primordial Star}
\author{Tom Abel, John H. Wise}
\affil{Kavli Institute for Particle Astrophysics and Cosmology, Stanford University, 2575 Sand Hill Road, Menlo Park, CA 94025}
\email{tabel@stanford.edu}
\and
\author{Greg L. Bryan}
\affil{Department of Astronomy, Columbia University, 1328 Pupin Physics Lab, 550 West 120th Street, New York, NY 10027}

\begin{abstract}
The concordance model of cosmology and structure formation predicts the formation of isolated very massive stars at high redshifts in dark matter dominated halos of $10^5$ to $10^6\Msun$. These stars photo-ionize their host primordial molecular clouds, expelling all the baryons from their halos. When the stars die, a relic \ion{H}{2} region is formed within which large amounts of molecular hydrogen form which will allow the gas to cool efficiently when gravity assembles it into larger dark matter halos. The filaments surrounding the first star hosting halo are largely shielded and provide the pathway for gas to stream into the halo when the star has died. We present the first fully three dimensional cosmological radiation hydrodynamical simulations that follow all these effects. A novel adaptive ray casting technique incorporates the time dependent radiative transfer around point sources. This approach is fast enough so that radiation transport, kinetic rate equations, and hydrodynamics are solved self-consistently. It retains the time derivative of the transfer equation and is explicitly photon conserving. This method is integrated with the cosmological adaptive mesh refinement code {\sl enzo}\, and runs on distributed and shared memory parallel architectures. Where applicable the three dimensional calculation not only confirm expectations from earlier one dimensional results but also illustrate the multi--fold hydrodynamic complexities of \ion{H}{2} regions. In the absence of stellar winds the circumstellar environments of the first supernovae and putative early gamma--ray bursts will be of low density $\sim 1$ cm$^{-3}$. Albeit marginally resolved, ionization front instabilities lead to cometary and elephant trunk like small scale structures reminiscent of nearby star forming regions. 
\end{abstract}
\keywords{stars: formation; ISM: \ion{H}{2} regions; cosmology: theory}
\submitted{April 9th, 2006}
\maketitle

\section{Motivation}\footnote{Visualizations and animations of the simulations presented here can be found at {\tt http://www.slac.stanford.edu/} {\tt $\sim$jwise/research/RT1/}}
With the fundamental cosmological parameters pinned down by recent observations, questions related to first structure formation have now, in principle, no free parameters. Hydrodynamical simulations that start with cosmological initial conditions have shed light on the nature of the first luminous objects in the universe \markcite{1998ApJ...508..518A, 1999AIPC..470...58N, 2000ApJ...540...39A, 2002Sci...295...93A, 2002MNRAS.330..927H, 2003ApJ...592..645Y}({Abel} {et~al.} 1998; {Norman}, {Abel}, \&  {Bryan} 1999; {Abel}, {Bryan}, \&  {Norman} 2000, 2002; {Hutchings} {et~al.} 2002; {Yoshida} {et~al.} 2003). \markcite{2002Sci...295...93A}{Abel} {et~al.} (2002) found that the first luminous objects are isolated massive stars with masses somewhere in the range between 30 and 300 solar masses depending on how much feedback from the proto-star affects the accretion rate. The stellar build up from predicted accretion rates in the absence of feedback have been confirmed in smooth particle hydrodynamic simulations~\markcite{2004NewA....9..353B}({Bromm} \& {Loeb} 2004). There are physical reasons from proto-stellar evolution that accretion may come to an end when the star reaches $\sim 100\Msun$ \markcite{2003ApJ...589..677O}({Omukai} \& {Palla} 2003) but full radiation hydrodynamical simulations, even in one dimension, have not been possible to follow primordial stars up to the zero age main sequence. The early proto-stellar evolution, however, is understood in some detail \markcite{1998ApJ...508..141O, 2002MNRAS.334..401R, 2004MNRAS.348.1019R}({Omukai} \& {Nishi} 1998; {Ripamonti} {et~al.} 2002; {Ripamonti} \& {Abel} 2004). 

The immediate relevance of primordial massive stars to cosmological reionization has prompted a number of investigations even though radiative transfer effects could only be accounted for crudely \markcite{2000ApJ...535..530G, 2001NewA....6..437G, 2004MNRAS.350...47S, 2005ApJ...628L...5O, 2006ApJ...639..621A}({Gnedin} 2000; {Gnedin} \& {Abel} 2001; {Sokasian} {et~al.} 2004; {O'Shea} {et~al.} 2005; {Alvarez}, {Bromm}, \&  {Shapiro} 2006) or in non-cosmological two-dimensional calculations~\markcite{2004MNRAS.348..753S}({Shapiro}, {Iliev}, \&  {Raga} 2004). In spherical symmetry, the evolution of the first \ion{H}{2} regions was studied by \markcite{2004ApJ...610...14W}{Whalen}, {Abel}, \&  {Norman} (2004) and \markcite{2004ApJ...613..631K}{Kitayama} {et~al.} (2004). These authors found that when the ionization front slows to become D-type it accelerates all the baryonic material to ten times the escape velocity of the dark matter halos that host these first stars. However, neither the stability of the ionization front, nor the impact of photo-ionization on the surrounding filaments could be addressed in the simplified one dimensional models. Significant amounts of molecular hydrogen are formed in relic \ion{H}{2} regions facilitated by their large initial electron fractions \markcite{2002ApJ...575...33R, 2005ApJ...628L...5O}(e.g. {Ricotti}, {Gnedin}, \&  {Shull} 2002; {O'Shea} {et~al.} 2005). Whether, photo-ionization leads to a net increase, decrease, or a delay in star formation remains controversial \markcite{1996ApJ...467..522H, 2000ApJ...534...11H, 2001ApJ...551..599H, 2000MNRAS.314..611C, 2003MNRAS.346..456O, 2005ApJ...628L...5O, 2006ApJ...639..621A}({Haiman}, {Rees}, \&  {Loeb} 1996; {Haiman}, {Abel}, \&  {Rees} 2000; {Haiman}, {Abel}, \&  {Madau} 2001; {Ciardi} {et~al.} 2000; {Oh} \& {Haiman} 2003; {O'Shea} {et~al.} 2005; {Alvarez} {et~al.} 2006). These outstanding questions will be resolved by direct numerical simulations that accurately follow the radiation transport as well as the cosmological hydrodynamics. Such calculations are presented in this {\sl Letter}\ for the first time.


\section{Simulations}

The simulations presented here implement an adaptive ray casting scheme into the cosmological adaptive mesh refinement hydrodynamics code {\sl enzo}\,~\markcite{1997ASPC..123..363B, 1999numa.conf...19N, Bryan:2001}({Bryan} \& {Norman} 1997; {Norman} \& {Bryan} 1999; {Bryan}, {Abel}, \& {Norman} 2001). We first describe this new method and then give details on the simulation setup and the implementation of star formation and assumptions about the formed population III stars. 
 
\subsection{Radiative Transfer}
 \markcite{1998MmSAI..69..455N}{Norman}, {Paschos}, \&  {Abel} (1998) have argued that radiative transfer for simulations of cosmological reionization would do best to split the highly anisotropic point sources from the diffuse sources, such as recombinations in high density regions. For the point sources, ray tracing is an effective way of following accurately the evolution of \ion{H}{2} regions around few sources \markcite{1999ApJ...523...66A, 2001NewA....6..359S}({Abel}, {Norman}, \&  {Madau} 1999; {Sokasian}, {Abel}, \&  {Hernquist} 2001). However, the many millions of rays required add unnecessary work when they are traced repeatedly through grid cells close to the radiation sources. Adaptive ray tracing offers an efficient alternative~\markcite{2002MNRAS.330L..53A}({Abel} \& {Wandelt} 2002) which has been successfully applied for post processing of cosmological simulations \markcite{2002ApJ...572..695R, 2002MNRAS.332..601S, 2003MNRAS.344..607S}({Razoumov} {et~al.} 2002; {Sokasian}, {Abel}, \&  {Hernquist} 2002; {Sokasian} {et~al.} 2003). Here we incorporate it self-consistently with the hydrodynamics.

For each radiation source at every fixed radiative time step we generate 768 (level three in the HEALPix pixelization of \markcite{2005ApJ...622..759G}{G{\'o}rski} {et~al.} (2005)) photon packages for each of four implemented frequency bins. The simulations presented here use only one of these bins but treat two types of radiation fields. We use a bin centered on $28.4$\, eV appropriate for hydrogen ionizing photons from a $100\Msun$ primordial star. The second radiation type is an optically thin $1/r^2$ flux for the Lyman Werner band dissociating photons is constrained to distances shorter than the light travel time.  The chemical abundances of the nine species model (\ion{H}{1}, \ion{H}{2}, \ion{He}{1}, \ion{He}{2}, \ion{He}{3}, e$^-$, H$^-$, H$_2^+$, H$_2$) of  \markcite{1997NewA....2..181A}{Abel} {et~al.} (1997) are solved for. The transfer and non-equilibrium chemistry and energy solver \markcite{1997NewA....2..209A, 1997NewA....2..181A}({Anninos} {et~al.} 1997; {Abel} {et~al.} 1997) sub-cycle on every grid. The photon packages are traced at the speed of light over constant time-steps of $170$\, yrs. while the hydrodynamics continues to use the usual adaptive hydro time steps. In the language of \markcite{2002MNRAS.330L..53A}{Abel} \& {Wandelt} (2002) we use 5.1 rays per local grid cell resolution. I.e. when the area associated with a ray becomes larger than 20\% of $dx^2$ of the local grid cell the photon package is discontinued and four child rays drawn from the next level of the HEALPix hierarchy are created sharing the photons of the parent ray. This approach effectively retains the time derivative of the radiative transfer equation and straightforwardly incorporates cosmological redshifting. We treat recombination to the ground state photons by the Case B approximation \markcite{1989agna.book.....O}({Osterbrock} 1989). In these first simulations presented here we neglect radiation pressure which only is marginally important in the early stages of the evolution of these \ion{H}{2} regions \markcite{1995MNRAS.273..249H,2004ApJ...613..631K}({Haehnelt} 1995; {Kitayama} {et~al.} 2004). We have verified that our implementation correctly reproduced analytical solutions of ionization front propagation using the test cases described in  \markcite{1999ApJ...523...66A}{Abel}, {Norman}, \&  {Madau} (1999). The coupled radiation hydrodynamics reproduce 1--D code tests of \markcite{2004ApJ...610...14W}{Whalen} {et~al.} (2004) in 3--D.

\subsection{Simulation Setup}
We present here calculations that have a top grid with $128^3$ grid resolution within a $250$\, comoving kpc volume with periodic boundary conditions. A subgrid at twice the topgrid resolution with $92^3$ grid cells is centered on the Lagrangian volume of a halo identified in earlier dark matter only runs. This affords an effective $256^3$ resolution with 30 solar mass dark matter particles in the high resolution region. The initial conditions, corresponding to an initial redshift of 130, are made with {\sl grafic} \markcite{2001ApJS..137....1B}({Bertschinger} 2001) for a cosmology with $(\Omega_B\,h^2, \ \Omega_M, \ h, \ \sigma_8, \ n)=(0.024,\ 0.27,\ 0.72,\ 0.9,\ 1)$ where the constants have the usual meaning of an inflationary Friedman--Robertson--Walker cosmology as e.g. in \markcite{2003ApJS..148..175S}{Spergel} {et~al.} (2003). The code refines when the baryon density exceeds four or the dark matter density exceeds eight times their initial almost uniform values. Additionally, the code refines to ensure that the local baryonic Jeans length is always resolved with at least 16 cells.

\subsection{Assumptions about the First Stars}
Stars are created automatically in regions that have an overdensity greater than $5 \times 10^5$ and a molecular hydrogen mass fraction larger than $5 \times 10^{-4}$.  \markcite{2002Sci...295...93A}{Abel} {et~al.} (2002) showed that when this criteria is fulfilled, first star formation occurs within 10 Myr.  We assume that the resulting star has a mass of 100 $M_\odot$ and shines for 2.7 Myr with $1.23 \times 10^{50}$ hydrogen ionizing photons per second, as in the no mass loss model of Schaerer (2002). These stellar particles are created by taking 50\% of the mass in grid cells within a radius that contains twice the stellar mass which is an input parameter. At the end of the star's life we continue to track the particle and continue to include it in the gravitational potential calculation. This mimics the formation of a black hole without a supernovae. The final states of these stars are remarkably uncertain \markcite{2002ApJ...567..532H}({Heger} \& {Woosley} 2002). We focus on this complete black hole formation scenario to separate the effects of the main sequence radiation from the effects of the explosions and metal mixing which will be the subject of a forthcoming publication.  

\section{Results}\label{results}

We present first the time evolution by examining radial profiles at several output times and then describe the complex morphologies using  projections through parts of the simulation volume. 

\subsection{Time Evolution}

The evolution of the dark matter merging and gas dynamics at redshifts below 130 are typical of previous simulations \markcite{2002Sci...295...93A, 2001ApJ...548...68J, 2003ApJ...592..645Y}({Abel} {et~al.} 2002; {Jang-Condell} \& {Hernquist} 2001; {Yoshida} {et~al.} 2003). In this realization, the first star forms during an ongoing merger of two dark matter halos. The larger one ($M_{tot}=5.5\times 10^{5}\Msun$) had other recent mergers and shows more turbulence. The star forms at redshift $z= 20.4$ in the smaller ($M_{tot}=4.0\times 10^{5}\Msun$) of the two. It moves at $0.2$, 3.6, and 3.1 km/s vs. the star forming halo, the center of mass of the merging object and the cosmic microwave background, respectively. 

\begin{figure}
\vspace{-0.5cm}
\centerline{\resizebox{\columnwidth}{!}{\includegraphics[width=\columnwidth]{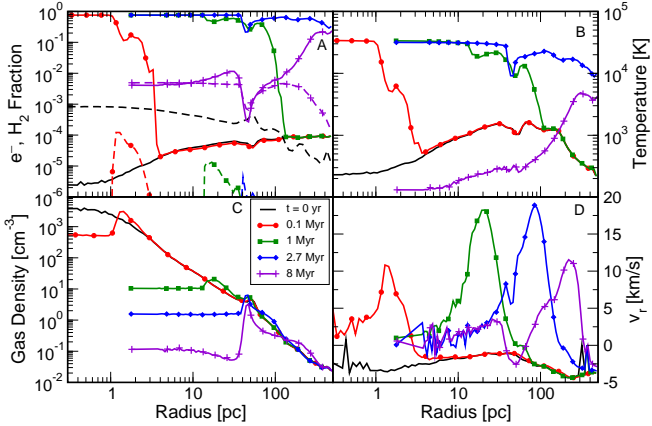}}\vspace{0.3cm}}
\caption{Mass weighted radial profiles around the position of the star that self-consistently formed in this cosmological simulation. A. shows both the electron number fraction (solid) and H$_2$ mass fractions (dashed) for 5 different output times, 0, 0.1, 1, 2.7, and 8 Myrs after the star is born. B. gives temperature profiles. Panels~C shows the density evolution.
Panel~D gives the radial velocity.  }
\label{rp}
\vspace{-0.1cm}
\end{figure}

Figure~\ref{rp} gives mass weighted spherically averaged quantities as distance from the location of the star. Panel~\ref{rp}A gives the time evolution of the ionized fraction of the gas together with the H$_2$ fraction (dashed). Within  100,000 years the ionized region expanded to one parsec in size, is D-type and is driving a shock ahead of it as can be seen in the temperature~(panel~\ref{rp}B) and density evolution~(\ref{rp}C). By the time the star switches off at 2.7 Myrs the spherically averaged mass weighted radial velocity has a maximum of $\sim 20$ km$/$s. This coherent outflow leads to central densities $\sim 1$ cm$^{-3}$ and has created a shell of material at 50 parsecs from the star's initial position. Even after the star's death this shell continues to expand, lowering the central densities further and evacuating the largely unaffected dark matter halo. Qualitatively these results are in very good agreement with the one dimensional models of \markcite{2004ApJ...610...14W}{Whalen} {et~al.} (2004) and \markcite{2004ApJ...613..631K}{Kitayama} {et~al.} (2004). The R--type front breaks out of the halo at approximately $1.2$ Myr after the star formed. With a constant luminosity for the star and a lifetime of 2.7 Myr this translates to an escape fraction of ionizing photons to the IGM of $\sim 56$\%.
The profiles at 8 Myrs, i.e. 5.3 Myr after the star switched off, show that very large amounts of molecular hydrogen ($f_{H_2} \sim 10^{-2}$) are being formed in the relic \ion{H}{2} region. However all that material is evacuated from the halo and the molecules find themselves at very low densities. The swept up shell is not Jeans unstable and is dispersed. At this time, 100,000 solar masses of gas are moving with 10 km/s outside their dark matter halo. 

A dense clump survived photo-evaporation at a distance of 50 parsecs from the star. From the radial velocities one can see that this clump is still infalling with respect to the position of the star. This clump is the center of a halo which started to merge with the star forming halo as the star was formed. At 8 Myrs it has a smaller molecular hydrogen fraction in its core than its surrounding. However, only 450 thousand years later the H$_2$ fraction grows and it cools to form a second star. This creates a second \ion{H}{2} region inside the one formed around the first star and drives yet more complex hydrodynamical flow.

\begin{figure*}
\centerline{\resizebox{6in}{!}{\includegraphics[width=6in]{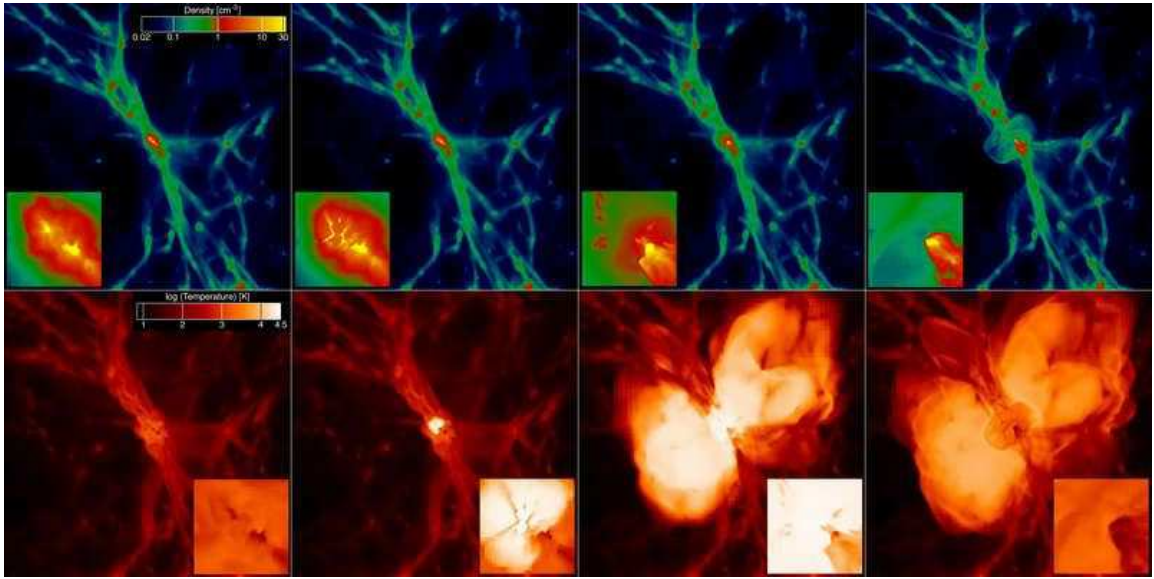}}}
\caption{Projections of a 64th of the simulation volume with 63 comoving kpc ($\sim 3$ proper kpc) side length centered on the position of the Pop III star. The rows give the integrated density squared  weighted gas density and temperature, respectively. From left to right the depicted times correspond to 0, 1, 2.7 and  8 Myrs after the star formed. The insets correspond to the same times, have the same color scale and show the central 150 parsec. No smoothing has been applied to these images.}\label{proj}
\vspace{0cm}
\end{figure*}

\subsection{Three Dimensional Structure}

From the insets in Figure~\ref{proj} it is apparent how the I-type front heats and expels nearly all the gas in the halo.  In doing so, it also forms new density enhancements which are all destroyed by the end of the star's life. The most massive clump, in the lower-right of the inset, survives, and is the dense core of a merging halo.  The gas which is expelled forms a shock which continues to expand until it can be recaptured by a larger dark matter halo forming in this region. The insets are 150 parsec across and illustrate marginally resolved, complex small scale structure reminiscent of the cometary and Elephant trunk like shapes as discussed extensively in the inter-stellar medium literature since \markcite{1958RvMP...30.1053P}{Pottasch} (1958). 
These instabilities are generated during the D-type phase and are amplified and stretched after the I-front breaks out into the more rarified gas behind the swept-up shell.  These clumps and the shielded regions behind them are compressed by the hot, ionized gas into long thin structures that point towards the ionizing star.  The gas inside these structures has temperatures around a few thousand degrees and densities of order a few particles per $cm^{-3}$.  By the end of the star's life, they have been completely dispersed.

The filaments along which the star forming halo forms are well shielded, leading to a butterfly shape for the ionized region similar to the ones first shown in~\markcite{1999ApJ...523...66A}{Abel} {et~al.} (1999).  The density structure in the filaments visible in these projections remain largely unaffected.  The ionized region has been heated to $\sim 3\times 10^4$ K and drives a shock into the intergalactic medium long after the star died, although these shock fronts move much more slowly than the original I-front speed, and will eventually stall as the relic region cools. The larger scale cosmic web changes hardly at all during the short lifetime of the star.

We will return to the longer-term impact of this ionization region on subsequent structure formation in a future paper, but already it is clear that the ionization preferentially heats low density gas and leaves high-density gas relatively untouched (at least the gas outside the immediate halo of the first star).  This will lead to a smaller entropy increase than predicted in simple models (e.g., Oh \& Haiman 2003), and so may blunt the predicted negative feedback from such relic \ion{H}{2} regions.

\section{Conclusions}\label{conc}

Early \ion{H}{2} regions have a profound impact on the earliest structures. They displace a large amount of baryons, raise the gas entropy, and enable the formation of large amounts of molecular hydrogen after the star dies. If Pop III stars indeed have little mass loss before they die \markcite{2001ApJ...550..890B}({Baraffe}, {Heger}, \&  {Woosley} 2001) our results illustrate that putative gamma ray bursts and supernovae will occur in a low density environment with densities $\sim 1$ cm$^{-3}$ \markcite{2004ApJ...604..508G}({Gou} {et~al.} 2004). The mixing of the first heavy elements occurs outside of dark matter halos in the intergalactic medium in regions that eventually will re-collapse to form galaxies. Consequently, the elements necessary for life may have been spread early and widely. 

The rich subject of ionization front instabilities \markcite{1979ApJ...233..280G,1996ApJ...469..171G}(cf. {Giuliani} 1979; {Garcia-Segura} \& {Franco} 1996) is important in early as well as galactic star formation. It can now be addressed in three dimensions at high spatial resolution using our radiation hydrodynamical adaptive mesh simulation techniques.

\acknowledgments{This work was supported by NSF CAREER award AST-0239709 from the National Science Foundation. GB acknowledges support through NSF grants AST-0507161 and AST-0547823. The simulation described here was carried out using eight processors of the SGI Altix system at KIPAC for one day. We gratefully acknowledge the SLAC computing services department for their administering and maintaining this great resource.}




\end{document}